\documentclass[prb,twocolumn,showpacs,preprintnumbers,superscriptaddress,amsmath,amssymb]{revtex4}

\usepackage{graphicx}
\usepackage{dcolumn}

\begin{document}

\title{Phase diagram of the PrFeAsO$_{1-x}$F$_{x}$ superconductor}


\author{C. R. Rotundu}
\affiliation{Materials Sciences Division, Lawrence Berkeley National
Laboratory, Berkeley, CA 94720, USA}
\author{D. T. Keane}
\affiliation{DND-CAT Advanced Photon Source, Argonne National
Laboratory, Argonne, IL 60439, USA}
\author{B. Freelon}
\affiliation{Department of Physics, University of California,
Berkeley, CA 94720, USA}
\author{S. D. Wilson}
\affiliation{Materials Sciences Division, Lawrence Berkeley National
Laboratory, Berkeley, CA 94720, USA}
\author{\\*A. Kim}
\affiliation{Department of Physics, University of California,
Berkeley, CA 94720, USA}
\author{P. N. Valdivia}
\affiliation{Department of Materials Science and Engineering, University of California,
Berkeley, CA 94720, USA}
\author{E. Bourret-Courchesne}
\affiliation{Life Sciences Division, Lawrence Berkeley National
Laboratory, Berkeley, CA 94720, USA}
\author{R. J. Birgeneau}
\affiliation{Materials Sciences Division,
Lawrence Berkeley National Laboratory, Berkeley, CA 94720, USA}
\affiliation{Department of Physics, University of California,
Berkeley, CA 94720, USA}
\affiliation{Department of Materials Science and Engineering, University of California,
Berkeley, CA 94720, USA}
\date{\today}

\begin{abstract}

The electronic phase diagram of PrFeAsO$_{1-x}$F$_{x}$ (0$\leq$x$\leq$0.225) has been determined using synchrotron X-ray powder diffraction,
magnetization and resistivity measurements. The structural transition temperature is suppressed from 154 K to $\approx$120 K and the magnetic phase
transitions of both iron and praseodymium ions are completely suppressed by x$\approx$0.08 fluorine doping, coinciding with the emergence of
superconductivity. The optimal doping is x$\approx$0.15 when T$_{C}$=47 K, while the maximum solubility of fluorine in PrFeAsO$_{1-x}$F$_{x}$ is reached around x=0.22. The structural, magnetic and superconducting phase diagram is presented.

\end{abstract}


\maketitle

\section{Introduction}

The discovery of superconductivity in the oxypnictides \emph{Ln}\emph{M}\emph{Pn}O (\emph{Ln}=La-Nd, Sm, Gd; \emph{M}=Fe, Co, Ni, Ru; \emph{Pn}=P and As)
\cite{Kamihara,GFChen,XHChen,Ren1} with a ZrCuSiAs-type structure \cite{Zimmer,Quebe} sets a milestone in the field of superconductivity.
The first step in the study of a newly discovered superconductor and towards elucidation of the nature of superconductivity itself is the
determination of the phase diagram. It is natural then to compare the pnictide phase diagram with that of the only other class of high temperature superconductors,
the cuprates \cite{Zhao1,Kivelson,Uemura,Hess}. As in the cuprates, superconductivity arises when a so-called parent
non-superconducting compound is doped with charge carriers. It has been demonstrated that the parent compound, \emph{Ln}FeAsO, can be doped with holes when
\emph{Ln}$^{3+}$ is replaced partially by a divalent ion (La$_{1-x}$Sr$_{x}$FeAsO \cite{HHWen}, Pr$_{1-x}$Sr$_{x}$FeAsO \cite{Mu}, Sr$^{2+}$). Correspondingly, $n$-type doping is realized
either by substitution of \emph{Ln}$^{3+}$ by a tetravalent ion (Gd$_{1-x}$Th$_{x}$FeAsO \cite{Wang}, Th$^{4+}$), or partially replacing
O$^{2-}$ by F$^{-}$. Among the iron-pnictides, to-date, the electron-doped (O$_{1-x}$F$_{x}$) iron-arsenides \emph{Ln}FeAsO$_{1-x}$F$_{x}$ have the highest
T$_{C}$'s reported. Until now, there have been reports on the phase diagrams of \emph{Ln}FeAsO$_{1-x}$F$_{x}$,
\emph{Ln}=La \cite{Kamihara,Kohama,Luetkens,Klingeler,Huang,Hess}, Ce \cite{GFChen,Zhao1}, Nd \cite{GFChen2}, and Sm \cite{Liu,Margadonna,Drew}.
Within the phase diagram of CeFeAsO$_{1-x}$F$_{x}$, J. Zhao $et$ $al.$ \cite{Zhao1} argue
for a suppression of the antiferromagnetism (AFM) with doping such that the magnetic order vanishes in close proximity to the superconductivity. This has been confirmed
in other \emph{Ln}FeAsO$_{1-x}$F$_{x}$ systems \cite{GFChen,GFChen2,Liu,Zhao1,Kohama,Luetkens,Klingeler,Huang,Hess}. Yet, in SmFeAsO$_{1-x}$F$_{x}$
there are reports of coexistence of static AFM order with superconductivity \cite{Margadonna,Drew}. Also, a signature of the Sm$^{3+}$ ions\textquoteright\ magnetism has been
reported in optimal superconducting SmFeAsO$_{0.85}$F$_{0.15}$ \cite{Ding}.
More recent studies on homologue systems \cite{Luetkens,Hess,Klingeler} report a rather \emph{abrupt}\ (first-order-like) change in the structural
and magnetic order parameters at the boundary of superconductivity. Despite wide efforts toward a unified picture, a consensus has not yet been reached.
Without any doubt, the correctness of the phase diagrams relies on an accurate determination of the fluorine ion concentration \cite{Hess,Kohler}. In this paper, we use synchrotron X-ray scattering, magnetization and resistivity measurements to map the
structural, magnetic and superconducting phase diagram of the less studied PrFeAsO$_{1-x}$F$_{x}$.

\section{Experimental Procedure}

PrFeAsO$_{1-x}$F$_{x}$ polycrystalline samples were synthesized through a two-step standard high temperature solid state chemical reaction using
stoichiometric amounts of PrAs, Fe (4N8), Fe$_{2}$O$_{3}$ (5N), and PrF$_{3}$ (4N) as starting materials \cite{XHChen,Liu}.
The PrAs binary used in the final reaction was synthesized by reacting Pr (3N) and As (4N) powders.
The Pr:As=1:1 mix was encapsulated within a tantalum tube that was sealed inside a quartz tube
under a low pressure Ar atmosphere, slowly heated to 500$^\circ$ C, and held at that temperature for 5 h. The powder was ground,
mixed, re-sealed, and heated again to 900$^\circ$ C for 10 h. The stoichiometric constituents were then thoroughly ground, mixed, and finally pressed
into pellets using a cold isostatic press  with a pressure of 0.38 GPa. To avoid direct contact with the quartz tube and possible Si contamination, the
pellets were wrapped in tantalum foil before being sealed under a low atmosphere of Ar gas in quartz tubes. They were heated to 1150$^\circ$ C for 50 h.
All preparatory steps except the annealing were performed in a glovebox under a high purity Ar gas atmosphere. The single phase
was checked after each step using XRD with Cu K$\alpha$ radiation at room temperature.

A discrepancy between the nominal and real fluorine content has been
systematically reported in homologue systems \cite{Hess,Kohler}. Therefore we determined the actual fluorine concentration by Wavelength Dispersive
X-ray Spectroscopy (WDS) for each concentration. Therefore, all fluorine concentrations reported in this paper are the as-measured values and not the
nominal starting compositions.
Resistivity, dc- and ac-susceptibility measurements down to 1.8 K were performed using a Quantum Design Physical Property Measurements System.
Electrical contacts with the samples for resistivity measurements were made using Pelco colloidal silver paste through attaching thin gold wires in a four-probe
configuration. The silver paste was cured at 100$^\circ$ C for up to 30 min in  order to avoid sample degradation. The excitation current of 2 mA
was optimized for the best signal-to-noise ratio and to prevent overheating of the samples by checking the linear I-V characteristic.
Because the absolute values of the resistivity data carry an uncertainty of up to 30\%\ due to the slightly off-rectangular
shape of the bar-shaped samples and the size of the contact pads, we report instead resistivity data as normalized to the room temperature
values. The ac susceptibility was measured with a 10 Oe modulated field of frequency f=1 kHz, while dc measurements were carried out in 200 Oe.
Synchrotron X-ray powder diffraction measurements with an incident-beam wavelength $\lambda$=0.31 {\AA} were performed at DND-CAT, Advanced Photon Source, Argonne National Laboratory.

\section{Results and Discussion}
The parent PrFeAsO antiferromagnetic semimetal crystalizes in a $\emph{P4/nmm}$ tetragonal structure at high temperatures but undergoes a $\emph(Cmma)$
orthorhombic transformation at low temperatures \cite{Zhao2}, which is characteristic of all \emph{Ln}FeAsO systems.
In order to obtain insight into the evolution of the crystallographic structure with fluorine doping and its relevance for superconductivity, we performed
synchrotron X-ray powder diffraction measurements on the undoped parent PrFeAsO, two non superconducting samples x=0.059 and 0.078, and one
under-doped superconducting concentration with x=0.082.

\begin{figure}[h]
\begin{center}\leavevmode
\includegraphics[width=1.05\linewidth]{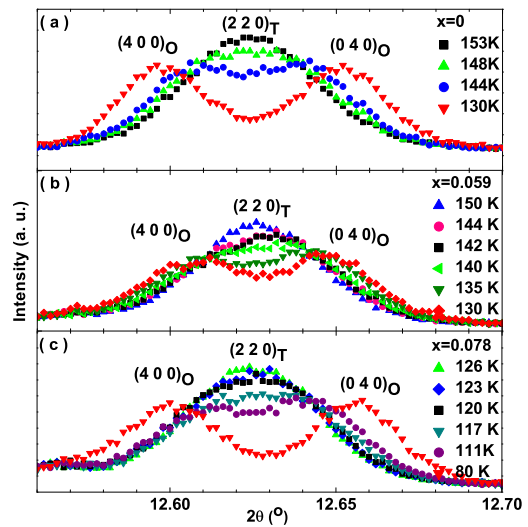}
\caption {The profile of the (2 2 0)$_{T}$ XRD peak of the $\emph{P4/nmm}$ tetragonal structure splitting bellow T$_{S}$
into the (4 0 0)$_{O}$ and (0 4 0)$_{O}$ reflections of the $\emph{Cmma}$ orthorhombic phase of non-superconducting PrFeAsO$_{1-x}$F$_{x}$ (a) x=0,
T$_{S}$=154 K, (b) x=0.059, T$_{S}$=142 K, and (c) x=0.078, T$_{S}$=120 K. ``\emph{T}'' and ``\emph{O}'' subscripts denote tetragonal and orthorhombic,
respectively.}\label{fig1}\end{center}\end{figure}

Figure 1 shows 2$\theta$ scans through the (2 2 0)$_{T}$ Bragg peak of the high
temperature $\emph{P4/nmm}$
tetragonal structure. The peak broadens and splits into the (4 0 0)$_{O}$ and (0 4 0)$_{O}$ peaks of the $\emph{Cmma}$ orthorhombic phase upon cooling.
``T'' and ``O'' subscripts denote tetragonal and orthorhombic, respectively. The corresponding structural transition temperatures found within
$\pm$2 K resolution are T$_{S}$=154 K for the parent, T$_{S}$=142 K for x=0.059, and T$_{S}$=120 K for
the fluorine doping preceding superconductivity x=0.078.

\begin{figure}[h]
\begin{center}\leavevmode
\includegraphics[width=1.05\linewidth]{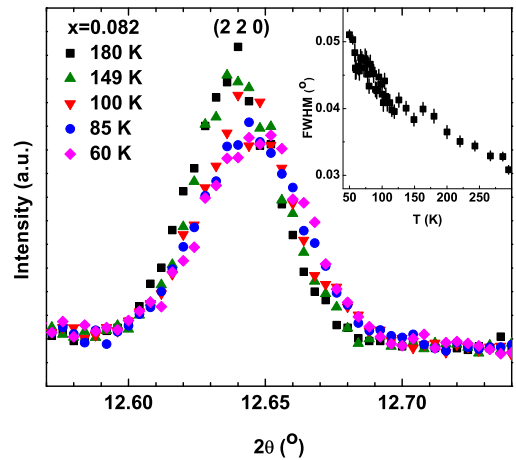}
\caption {The profile of the (2 2 0) XRD peak of the tetragonal superconducting PrFeAsO$_{1-x}$F$_{x}$, x=0.082. The inset shows the full-width at
half-maximum (FWHM) of the peak versus temperature.}\label{fig2}\end{center}\end{figure}

Figure 2 shows the evolution of the (2 2 0) Bragg peak for the superconducting x=0.082 sample
upon cooling to 50 K. The full-width at half-maximum (FWHM) of the peak is plotted versus temperature in the inset of Fig. 2 and shows a continuous evolution
down to 50 K. Upon cooling, we observe a continuous slight broadening of the (2 2 0) peak; yet we were not able to resolve any definite inflection
point in the
FWHM versus T indicative of a signature of the tetragonal to orthorhombic structural phase transition \cite{Zhao1}. Therefore, our data support a picture
of a rather abrupt suppression of the orthorhombic phase at the boundary of superconductivity. Our best estimate is a complete suppression of
T$_{S}$ from 120 K within 1.5$\%$ fluorine doping of the critical value for superconductivity ($x$ between 0.072 and 0.087).

\begin{figure}[h]
\begin{center}\leavevmode
\includegraphics[width=1.05\linewidth]{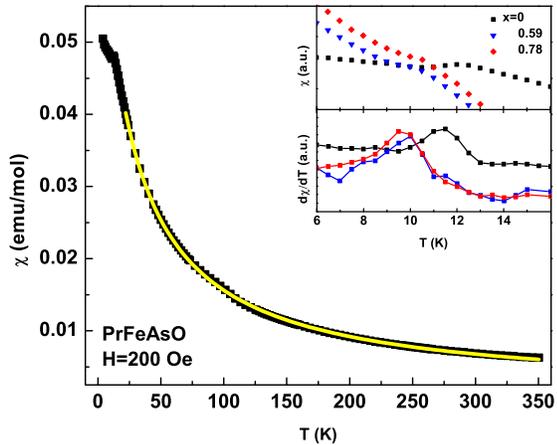}
\caption {The temperature-dependent magnetic susceptibility of PrFeAsO. The yellow (or light gray) curve is the result of a fit to Curie-Weiss law. The upper
inset shows the susceptibility of PrFeAsO$_{1-x}$F$_{x}$, with x=0, 0.059 and 0.078 around the Pr ion ordering temperature along with their
derivatives in the lower inset. In the lower panel the line is a guide-to-the-eye.}\label{fig3}\end{center}\end{figure}

Now we turn to our magnetic susceptibility data. Figure 3 shows the temperature dependent, dc-magnetic susceptibility of PrFeAsO measured in 200 Oe fitted
above the Pr ordering temperature with a Curie-Weiss law, $\chi(T)=\chi_{0}+C/(T+\theta)$, where $\chi_{0}$ is the temperature independent susceptibility,
$C=N\mu_{eff}^2/3k_{B}$ the Curie constant, $N$ is the number of Pr$^{3+}$ ions, and $\theta$ the Curie-Weiss temperature. The effective moment found
is $\mu_{eff}$=3.75$\mu_{B}$, slightly larger than the
isolated trivalent Pr ion value of 3.58$\mu_{B}$ \cite{Ashcroft}. The discrepancy could be due to the Pr$^{3+}$ CEF contribution to the susceptibility \cite{Wilson}.
It should be noted that the susceptibility data could be fitted with a Curie-Weiss law for \emph{Ln}FeAsO$_{1-x}$F$_{x}$, \emph{Ln}=Ce \cite{GFChen,McGuire},
Nd \cite{McGuire} and Gd \cite{Wang}, but not in the case of Ln=Sm \cite{Martinelli}.
The low temperature susceptibility is dominated by a sharp peak around 12 K in the pure PrFeAsO system, as well as in all other non-superconducting F$^{-}$ concentrations.
This is due to the AFM ordering (N\'{e}el temperature) of the Pr 4$\emph{f}$ electrons (Fig. 5(b)), consistent with previous reports from
neutron \cite{Zhao2}, magnetization \cite{McGuire}, and resistivity \cite{McGuire,Kimber} studies. The upper inset shows a moderated suppression of T$_{N}(Pr)$
with fluorine doping for three different concentrations x=0, 0.059 and 0.078 in PrFeAsO$_{1-x}$F$_{x}$ extending over the whole non-superconducting region.
In the lower inset the temperature derivative of the susceptibility around the ordering temperature is shown. Again, it should be noted that there is a weak to moderate
suppression of the ordering temperature with fluorine doping.

\begin{figure}[h]
\begin{center}\leavevmode
\includegraphics[width=1.05\linewidth]{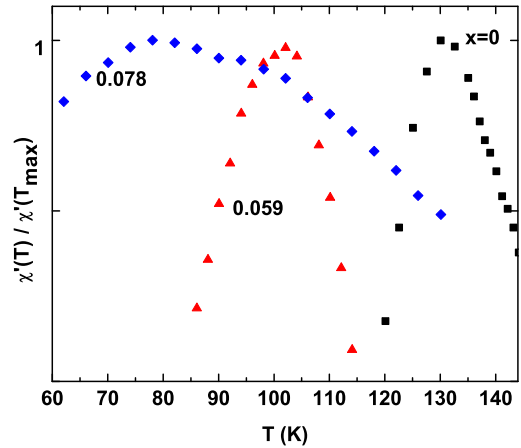}
\caption {Real component of ac susceptibility normalized to its
maximum of PrFeAsO$_{1-x}$F$_{x}$, with x=0, 0.059, and
0.078.}\label{fig4}\end{center}\end{figure}

Figure 4 presents the real components of the ac-susceptibilities normalized to their corresponding maxima for the same compositions (x=0, 0.059, 0.78)
discussed previously. The 130 K sharp peak in magnetization of the parent compound coincides with the magnetic ordering
temperature of the iron ion moments, as initially revealed by the neutron diffraction data \cite{Zhao2}. With fluorine doping, the magnetic ordering
temperature of the iron moments is continuously reduced from 130 K (x=0) to 80 K for x=0.078.

\begin{figure}[h]
\begin{center}\leavevmode
\includegraphics[width=1.09\linewidth]{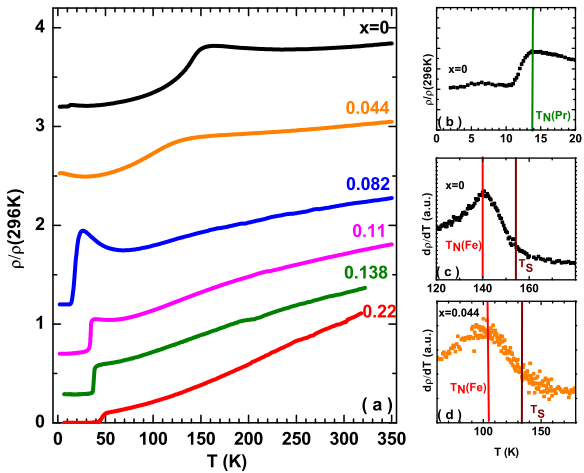}
\caption {(a) Resistivity of PrFeAsO$_{1-x}$F$_{x}$ for x=0, 0.044, 0.11, 0.138 and 0.225 divided by the room temperature value.
Data is shifted for clarity. (b) The resistivity near the Pr ion ordering temperature (T$_{N}$(Pr)) for the parent sample PrFeAsO.
The temperature derivative of the resistivity ($\partial\rho$/$\partial$$T$) for (c) x=0 and (d) x=0.044. T$_{S}$ and T$_{N}$ mark the structural and the magnetic
phase transitions, respectively.}\label{fig5}\end{center}\end{figure}

In Fig. 5 we show resistivity data for PrFeAsO$_{1-x}$F$_{x}$ with x=0, 0.044, 0.11, 0.138 and 0.225 divided by the room temperature value, $\rho(297$ $K)$.
The data are shifted for clarity. The behavior of $\rho(T)$ for the PrFeAsO compound is similar to that of the other \emph{Ln}FeAsO, \emph{Ln}=La, Ce,
Nd and Sm \cite{McGuire,Liu}. Upon cooling, the resistivity decreases showing metallic behavior. Around 150 K there is a broad peak associated with both the structural
phase transition from tetragonal to orthorhombic symmetry and the spin-density-wave (SDW) magnetic phase transition. Because of a smooth and continuous change in the physical
properties around these transitions, an indicator of the transition temperature is the temperature derivative of the resistivity
\cite{Klauss,Klingeler,McGuire,Hess}. Figures 5(c) and (d) show $\partial \rho/\partial T$ for the x=0 and 0.044 non-superconducting samples.
T$_{N}(Fe)$ and T$_{S}$ correspond to the two inflection points in $\partial \rho/\partial T$.
At around 13 K, the resistivity of the parent PrFeAsO exhibits a shoulder-like feature (Fig. 5(b)) that was attributed to the AFM ordering of the Pr ion
moments \cite{McGuire,Kimber}.

\begin{figure}[h]
\begin{center}\leavevmode
\includegraphics[width=0.8\linewidth]{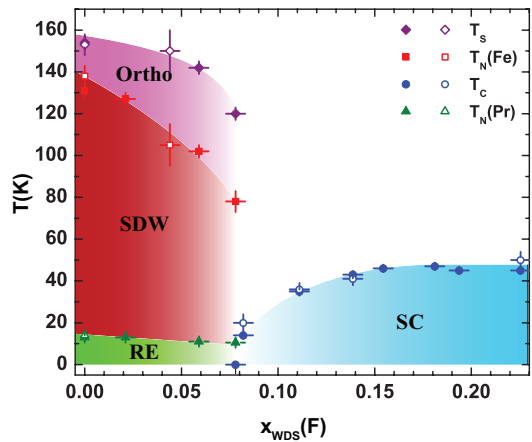}
\caption {The structural, magnetic and superconducting phase diagram of PrFeAsO$_{1-x}$F$_{x}$, 0$\leq$x$\leq$0.225 as determined from our synchrotron
X-ray powder diffraction, magnetization and resistivity measurements. The $P4/nmm$ to $Cmma$ phase transition as determined from X-ray powder diffraction
($\blacklozenge$) and from the temperature derivative of the resistivity $\partial\rho$/$\partial$$T$ ($\lozenge$). The N\'{e}el temperatures of Fe (T$_{N}$(Fe)) and
Pr (T$_{N}$(Pr)) ions as determined from magnetization ($\blacksquare$, $\blacktriangle$) and from $\rho$ or $\partial\rho$/$\partial$$T$ ($\square$, $\vartriangle$),
respectively. The superconducting transition temperatures T$_{C}$ for samples with fluorine doping between $\sim$0.08 and 0.225 were determined from
susceptibility ($\bullet$) and from the drop of the resistivity ($\circ$).}\label{fig6}\end{center}\end{figure}

Fig. 6 shows the phase diagram of PrFeAsO$_{1-x}$F$_{x}$ (0$\leq$x$\leq$0.225) as determined by the measurements reported in this paper.
Superconductivity appears after doping with charge carriers (electrons)
the parent AFM semimetal PrFeAsO. In this system, doping is realized by substituting O$^{2-}$ by F$^{-}$. Upon doping, the coupled
structural and AFM transitions are suppressed, in analogy with the behavior of the structural and magnetic transitions in La$_{2-x}$Sr$_{x}$CuO$_{4}$ \cite{Birgeneau}.
The phase diagram showing close proximity between AFM and SC resembles that of the electron-doped cuprates where the AFM persists right up to the SC dome.
The superconducting transition temperatures were determined by the onset of the diamagnetism in zero-field-cooled magnetization (data not shown)
or the drop of the resistivity. Optimal doping is achieved at a fluorine concentration of $\approx$0.15, in agreement with values reported for the
homologous systems CeFeAsO$_{1-x}$F$_{x}$ \cite{GFChen,Zhao1} and SmFeAsO$_{1-x}$F$_{x}$ \cite{XHChen,Liu,Drew} and as predicted by recent minimum
principle energy calculations \cite{XHuang}. \emph{Ln}FeAsO$_{1-\delta}$ oxygen deficient (optimally doped at $\delta$=0.15) with no fluorine doping obtained
by high pressure synthesis has been reported to be a superconductor \cite{Ren2}; this process is similar to the well known oxygenation of the cuprates.
It is worthwhile to mention that a maximum superconducting temperature transition around 47 K has been reported
previously for optimally doped samples synthesized by normal pressure solid state chemical reaction \cite{Bhoi2,Kimber}, while a slightly
enhanced T$_{C}$ of 52 K was reported for the samples synthesized using a high pressure method \cite{Ren1}. Our results show that the synthesis
of a sample with nominal fluorine concentration of 0.35 resulted in an actual fluorine x$_{WDS}$ substitution of 0.225, which represents the maximum
fluorine doping in PrFeAsO$_{1-x}$F$_{x}$. This value is consistent with recently reported WDS determined values in LaFeAsO$_{1-x}$F$_{x}$ and
SmFeAsO$_{1-x}$F$_{x}$ \cite{Kohler}
and the saturation of lattice parameters in LaFeAsO$_{1-x}$F$_{x}$ \cite{Kondrat}.

\section{Conclusions}

In summary, we have determined the complete structural, magnetic and superconducting phase diagram of PrFeAsO$_{1-x}$F$_{x}$ (0$\leq$x$\leq$0.22)
using synchrotron X-ray powder diffraction, magnetization and resistivity measurements.
We find a progressive suppression of the structural transition and magnetic ordering transitions of both iron and praseodymium ion moments
with increasing fluorine doping. In superconducting samples, near the edge of the
emergence of the superconductivity (x$\approx$0.08 fluorine doping), we were not able to detect any fraction of the orthogonal phase. The optimal doping was found to be at x$\sim$0.15 when T$_{C}$=47 K, and the maximum fluorine doping in PrFeAsO$_{1-x}$F$_{x}$ was reached around x=0.22 with superconductivity remaining robust.

The phase diagram is most similar to that of the hole-doped cuprates La$_{2-x}$Sr$_{x}$CuO$_{4}$ \cite{Birgeneau}. However, there are some important differences.  First, in La$_{2-x}$Sr$_{x}$CuO$_{4}$ there is a spin glass phase in between the 3D N\'{e}el antiferromagnet and the superconductor; the stripe-like spin fluctuations in the cuprate spin glass phase are rotated by 45 degrees, relative to those in the superconductor and the spin glass-superconductor transition at 0 K as a function of hole concentration is first order. Second, the tetragonal-orthorhombic structural transition temperature in La$_{2-x}$Sr$_{x}$CuO$_{4}$ varies smoothly across the insulating to superconductor boundary and, indeed, persists in $x$ beyond the value at which T$_{C}$ is a maximum.
By contrast, in the PrFeAsO$_{1-x}$F$_{x}$ system both the N\'{e}el order and the tetragonal-orthorhombic structural transition appear to vanish together quite rapidly, possibly in a first order way, as the fluorine concentration is varied through the critical value for superconductivity. In order to elucidate this further it will be necessary to prepare homogeneous samples in which the fluorine concentration, $x$, is varied in quite fine steps. This will represent a significant technical challenge. It will also be necessary to characterize the magnetism in samples in the transition region using both neutron scattering and a local technique such as muon spin resonance.
Of course, the ultimate goal is to determine which features are universal and which are details of a particular system. Specifically, one would like to determine whether or not the iron pnictides and the cuprates are in the same universality class or whether, in fact, the similarity in the phase diagrams is coincidental.

\begin{acknowledgments}
The authors thank to S. M. Hanrahan, C. Ramsey, K. Ross and J. Wu for technical support.
This work was supported by the Director, Office of Science,
Office of Basic Energy Sciences, U.S. Department
of Energy, under Contract No. DE-AC02-05CH11231
and Office of Basic Energy Sciences US DOE DE-AC03-
76SF008. Work at Advanced Photon
Source (APS) was performed at the DuPont-Northwestern-Dow
Collaborative Access Team (DND-CAT) located at Sector 5 APS. DND-CAT
is supported by E.I. DuPont de Nemours $\&$ Co., The Dow Chemical
Company and the State of Illinois. Use of the APS was supported by
the U. S. Department of Energy, Office of Science, Office of Basic
Energy Sciences, under Contract No. DE-AC02-06CH11357.

\end{acknowledgments}


\begin{thebibliography}{35}

\bibitem{Kamihara}
Y. Kamihara, T. Watanabe, M. Hirano, and H. Hosono, J. Am. Chem.
Soc. {\bf 130}, 3296 (2008)

\bibitem{GFChen}
G. F. Chen, Z. Li, D. Wu, G. Li, W. Z. Hu, J. Dong, P. Zheng, J. L. Luo, and N. L. Wang, Phys. Rev. Lett. {\bf 100},
247002 (2008).

\bibitem{XHChen}
X. H. Chen, T. Wu, G. Wu, R. H. Liu, H. Chen and D. F. Fang, Nature {\bf 453}, 761 (2008).

\bibitem{Ren1}
Z.-A. Ren, J. Yang, W. Lu, W. Yi, G.-C. Che, X.-L. Dong, L.-L. Sun, Z.-X.-Zhao,
Mat. Res. Innov. {\bf 12}, 105 (2008).

\bibitem{Zimmer}
B. I. Zimmer, W. Jeitschko, J. H. Albering, R. Glaum, and M. Reehuis, J. Alloys Compd. {\bf 229}, 238 (1995).

\bibitem{Quebe}
P. Quebe, L. J. Terb\"{u}chte, W. Jeitschko, J. Alloys Compd. {\bf 302}, 70 (2000).

\bibitem{Zhao1}
J. Zhao, Q. Huang, C. de la Cruz, S. Li, J. W. Lynn, Y. Chen, M. A.
Green, G. F. Chen, G. Li, Z. Li, J. L. Luo, N. L. Wang and P. Dai,
Nature Mater. {\bf 7}, 953 (2008).

\bibitem{Kivelson}
S. A. Kivelson and H. Yao, Nature Mater. {\bf 7}, 927 (2008).

\bibitem{Uemura}
Y. Uemura, Nature Mater. {\bf 8}, 253 (2009).

\bibitem{Hess}
C. Hess, A. Kondrat, A. Narduzzo, J. E. Hamann-Borrero, R. Klingeler, J. Werner, G. Behr, and B. B\"{u}chner,
arXiv:0811.1601 (2008).

\bibitem{HHWen}
H.-H. Wen, G. Mu, L. Fang, H. Yang and X. Zhu,
Europhys. Lett. {\bf 82}, 17009 (2008).

\bibitem{Mu}
G. Mu, B. Zeng, X. Zhu, F. Han, P. Cheng, B. Shen, and H.-H. Wen, Phys. Rev. B {\bf 79}, 104501 (2009).

\bibitem{Wang}
C. Wang, L. Li, S. Chi, Z. Zhu, Z. Ren, Y. Li, Y. Wang, X. Lin, Y. Luo, S. Jiang, X. Xu, G. Cao, and Z. Xu,
Europhys. Lett. {\bf 83}, 67006 (2008).

\bibitem{Kohama}
Y. Kohama, Y. Kamihara, M. Hirano, H. Kawaji, T. Atake, and H. Hosono, Phys. Rev. B {\bf 78}, 020512(R) (2008).

\bibitem{Luetkens}
H. Luetkens, H.-H. Klauss, M. Kraken, F. J. Litterst, T. Dellmann, R. Klingeler, C. Hess,
R. Khasanov, A. Amato, C. Baines, M. Kosmala, O. J. Schumann, M. Braden, J. Hamann-Borrero,
N. Leps, A. Kondrat, G. Behr, J.Werner and B. B\"{u}chner, Nature Mater. {\bf 8}, 305 (2009).

\bibitem{Klingeler}
R. Klingeler, N. Leps, I. Hellmann, A. Popa, C. Hess, A. Kondrat, J. Hamann-Borrero, G. Behr, V. Kataev, B. B\"{u}chner,
arXiv:0808.0708 (2008).

\bibitem{Huang}
Q. Huang, J. Zhao, J. W. Lynn, G. F. Chen, J. L. Luo, N. L. Wang, and P. Dai,
Phys. Rev. B {\bf 78}, 054529 (2008).

\bibitem{GFChen2}
G. F. Chen, Z. Li, D. Wu, J. Dong, G. Li, W. Z. Hu, P. Zheng, J. L. Luo, and N. L. Wang, Chin. Phys. Lett. {\bf 25},
2235 (2008).

\bibitem{Liu}
R. H. Liu, G. Wu, T. Wu, D. F. Fang, H. Chen, S. Y. Li, K. Liu, Y. L. Xie, X. F. Wang, R. L. Yang,
L. Ding, C. He, D. L. Feng, and X. H. Chen, Phys. Rev. Lett. {\bf 101}, 087001 (2008).

\bibitem{Margadonna}
S. Margadonna, Y. Takabayashi, M. T. McDonald, M. Brunelli, G. Wu, R. H. Liu, X. H. Chen, K. Prassides,
Phys. Rev. B {\bf 79}, 014503 (2009).

\bibitem{Drew}
A. J. Drew, Ch. Niedermayer, P. J. Baker, F. L. Pratt, S. J. Blundell, T. Lancaster, R. H. Liu,
G.Wu, X. H. Chen, I.Watanabe, V. K. Malik, A. Dubroka, M. R\"{o}ssle, K.W. Kim, C. Baines
and C. Bernhard, Nature Mater. {\bf 8}, 310 (2009).

\bibitem{Ding}
L. Ding, C. He, J. K. Dong, T. Wu, R. H. Liu, X. H. Chen, and S. Y. Li,
Phys. Rev. B {\bf 77}, 180510(R) (2008).

\bibitem{Kohler}
A. K\"{o}hler, G. Behr, arXiv:0906.0326 (2009).

\bibitem{Zhao2}
J. Zhao, Q. Huang, C. de la Cruz, J. W. Lynn, M. D. Lumsden, Z. A. Ren, J. Yang, X. Shen, X. Dong,
Z. Zhao, and P. Dai, Phys. Rev. B {\bf 78}, 132504 (2008).

\bibitem{Ashcroft}
N. W. Ashcroft and D. N. Mermin, Solid State Physics (Brooks Cole, 1976), ISBN 0030839939.

\bibitem{Wilson}
S. D. Wilson, C. R. Rotundu $\emph{et al.}$, in preparation.

\bibitem{McGuire}
M. A. McGuire, R. P. Hermann, A. S. Sefat, B. C. Sales, R. Jin, D.
Mandrus, F. Grandjean, and G. J. Long, New Journal of Physics {\bf 11}, 025011 (2009).

\bibitem{Martinelli}
A. Martinelli, M. Ferretti, P. Manfrinetti, A. Palenzona, M. Tropeano, M. R. Cimberle, C. Ferdeghini, R. Valle,
C. Bernini, M. Putti, and A. S. Siri, Supercond. Sci. Technol. {\bf 21}, 095017 (2008).

\bibitem{Kimber}
S. A. J. Kimber, D. N. Argyriou, F. Yokaichiya, K. Habicht, S. Gerischer, T. Hansen, T. Chatterji, R. Klingeler, C. Hess, G. Behr,
A. Kondrat, and B. B\"{u}chner, Phys. Rev. B {\bf 78}, 140503(R) (2008).

\bibitem{Klauss}
H. -H. Klauss, H. Luetkens, R. Klingeler, C. Hess, F. J. Litterst, M. Kraken, M. M. Korshunov, I. Eremin, S. -L. Drechsler,
R. Khasanov, A. Amato, J. Hamann-Borrero, N. Leps, A. Kondrat, G. Behr, J. Werner, B. Buchner, Phys. Rev. Lett. {\bf 101}, 077005 (2008).

\bibitem{Birgeneau}
R. J. Birgeneau, C. Stock, J. M. Tranquada, and K. Yamada, J. Phys. Soc. Jpn. {\bf 75}, 111003 (2006).

\bibitem{XHuang}
X. Huang, arXiv:0807.0899 (2008).

\bibitem{Ren2}
Z.-A. Ren, G.-C. Che, X.-L. Dong, J. Yang, W. Lu, W. Yi, X.-L. Shen, Z.-C. Li, L.-L. Sun, F. Zhou and Z.-X. Zhao,
Europhys. Lett. {\bf 83}, 17002 (2008).

\bibitem{Bhoi2}
D. Bhoi, P. Mandal and P. Choudhury, Supercond. Sci. Technol. {\bf 21}, 125021 (2008).

\bibitem{Kondrat}
A. Kondrat, J. E. Hamann-Borrero, N. Leps, M. Kosmala, O. Schumann, J. Werner, G. Behr, M. Braden, R. Klingeler, B. B\"{u}chner, C. Hess,
arXiv:0811.4436 (2008).

\end{thebibliography}
\end{document}